\documentclass[letterpaper,11pt]{article}
\usepackage{graphics,graphicx,epsfig,wrapfig}
\usepackage{longtable}  
\usepackage{amsmath,verbatim,amssymb,color,lscape}
\usepackage{kotex}
\usepackage{natbib}
\usepackage{lastpage}
\usepackage{multirow} 
\usepackage{authblk}
\usepackage{bm}
\usepackage[ruled,vlined]{algorithm2e}
\usepackage[normalem]{ulem}
\usepackage[export]{adjustbox}

\newcommand{\hide}[1]{}

\makeatletter

\usepackage[outerbars,color]{changebar}
\ifx\pdfoutput\undefined
\else\ifnum\pdfoutput>0
  \usepackage{pdfcolmk}
\fi\fi
\cbcolor{black}

\usepackage[margin=1.0in]{geometry}
\usepackage{tikz}
\usetikzlibrary{bayesnet}
\usetikzlibrary{fit,positioning}

\usepackage[ruled,vlined]{algorithm2e}
\usepackage{algorithmic}

\usepackage{xr}

\newfont{\rmm}{cmr10 at 11pt}
\rmm

\title{\textbf{Impacts of Innovation School System in Korea: \\ A Latent Space Item Response Model with Neyman-Scott Point Process}}

\author[1,2]{Seorim Yi}
\author[1]{Minkyu Kim}
\author[1,3]{Jaewoo Park}
\author[4]{Minjeong Jeon}
\author[1,3]{Ick Hoon Jin}

\affil[1]{Department of Statistics and Data Science, Yonsei University}
\affil[2]{Department of Statistics and Data Science, University of Texas}
\affil[3]{Department of Applied Statistics, Yonsei University}
\affil[4]{School of Education and Information Studies, University of California, Los Angeles}

\date{}

\begin{document}
\maketitle

\begin{abstract}
South Korea's educational system has faced criticism for its lack of focus on critical thinking and creativity, resulting in high levels of stress and anxiety among students. As part of the government's effort to improve the educational system, the innovation school system was introduced in 2009, which aims to develop students' creativity as well as their non-cognitive skills. To better understand the differences between innovation and regular school systems in South Korea, we propose a novel method that combines the latent space item response model (LSIRM) with the Neyman-Scott (NS) point process model. Our method accounts for the heterogeneity of items and students, captures relationships between respondents and items, and identifies item and student clusters that can provide a comprehensive understanding of students' behaviors/perceptions on non-cognitive outcomes. Our analysis reveals that students in the innovation school system show a higher sense of citizenship, while those in the regular school system tend to associate confidence in appearance with social ability. 
\textcolor{black}{
A comparison with exploratory item factor analysis highlights our method's advantages in terms of uncertainty quantification of the clustering process and more detailed and nuanced clustering results. Our method is made available to an existing \texttt{R} package, {\tt lsirm12pl} \citep{go2022lsirm12pl}.  
}
\end{abstract}


\section{Introduction}

The education system in South Korea, recognized for students' consistently high performance in international assessments,  
has been criticized for its excessive emphasis on rote memorization and test-taking, hindering imagination, self-achievement, and critical thinking \citep{Kim2004, Randall2013, ART001518899}. To address these criticisms, the Gyeonggi Province Office of Education introduced the innovation school system in 2009 \citep{Gyeonggi:2012, Gyeonggi:2020}, which focuses on democracy and ethics while prioritizing creativity and student-centered learning. In order to maintain and/or repair the innovation school system policy, it is crucial to evaluate its effectiveness in promoting desired student behaviors.

Past studies have used various conventional methods, such as ANOVA, t-tests, and item response theory (IRT), to compare the effectiveness of the innovation school system with the traditional Korean school system. For instance, \citet{KimInnovation2019} and \citet{203589} employed ANOVA and regression analysis to identify significant differences between innovation and traditional school systems. Similarly, \citet{kim2019comparative} used IRT to compare the academic achievement of students in innovation high schools and general high schools in South Korea. They found that students in innovation high schools showed higher academic achievement than students in general high schools, particularly in math and science. Additionally, \citet{kimjeonnam2017} used t-tests to analyze the effectiveness of innovation schools based on the comparison of `democratic school management', `curriculum', and `satisfaction with education' between regular and innovation schools. Likewise, \citet{Ryu2020} employed t-tests to compare  the performance of the innovation schools in Gyeongnam. 
Although these previous studies have provided some insight into the differences between the two school systems, a full picture of how the innovative school system really affects student lives might not have been thoroughly  shown due to the restrictive assumptions of those conventional methods used in these studies.


Recently, \cite{jin2022hierarchical} applied an innovative approach, referred to as the network item response model (NIRM; \citealp{jin2019doubly}), to analyze the item-level differences in student responses to mental well-being questionnaire items between innovation schools and regular schools. 
By leveraging techniques from social network models, \cite{jin2022hierarchical} regarded students' responses to the multi-item questionnaire items as networks between respondents and networks between items and represented their relationships in a common Euclidean space, where shorter distances indicate stronger relationship, i.e., higher likelihood of endorsing the items (therefore, showing the corresponding behaviors) for the respondents. 
This way, the estimated latent space from NIRM can reveal subtle differences at the item level in students' mental well-being between the two school systems. Note that this is a unique advantage of the NIRM approach, in the sense that most other conventional methods analyze scale-level  aggregated student well-being measures, washing away potential item-level differences. 
Despite the advantages, \cite{jin2022hierarchical}'s approach had several limitations: 1) relationships between respondents and items were not directly modeled; 
2) the model accommodated  a multilevel structure of the latent spaces between schools in a complicated way, resulting in a less intuitive interpretation of the school-level latent space; 3) the interpretation of the  latent space configuration was somewhat subjective.

In this study, we develop and apply a new statistical approach that can shed new light on how the two school systems differ in terms of students' mental well-being. Our approach is related to \cite{jin2022hierarchical}'s approach, but we advance their methodology as follows: First, by regarding student response data as a bipartite network, we directly model the relationships (or interactions) between respondents and items. We utilize the latent space item response model \citep[LSIRM;][]{jeon2021mapping} to represent the item-by-person relationships in a two-dimensional latent space, called an interaction map. On the map, distances indicate unobserved similarities and dissimilarities between items, between respondents, and between items and respondents in terms of item endorsement likelihoods. Thus, by evaluating the differences in the item configuration of the interaction map between the two school systems, for example, we can understand which particular behaviors the students are more or less likely to show in the two school systems. Furthermore, we apply the Neyman-Scott (NS) point process model  \citep{thomas1949generalization, neyman1952theory} to automate the interpretation of the latent space configuration from the LSIRM by modeling clustering of the latent positions. 
\textcolor{black}{
Our method supplies a distribution of the number of clusters as well as contour plots that help inform the location of cluster centers. Such uncertainty quantification is a distinct advantage of our method over standard clustering methods in psychometrics. 
}
With the proposed approach, we investigate the differences between the two school systems concerning students' non-cognitive behaviors at the individual item level, with a particular emphasis on emotional and moral behaviors.   
\textcolor{black}{We implemented the proposed method in {\tt R}  and made it available to the freely available package {\tt lsirm12pl} \citep{go2022lsirm12pl}. }


The remainder of this paper is organized as follows. Section 2 describes the data that we will analyze in this study. We describe how we match the student samples from the two school systems for a more accurate and fairer comparison. \textcolor{black}{We introduce our approach and provide preliminary results after fitting the LSIRM. }
In Section 3, we apply the proposed approach to the data and interpret the results. In Section 4, we conclude the paper with a summary and discussion.

\section{Methods}

\begin{figure}[htbp]
    \centering
    \includegraphics[width=0.9\linewidth]{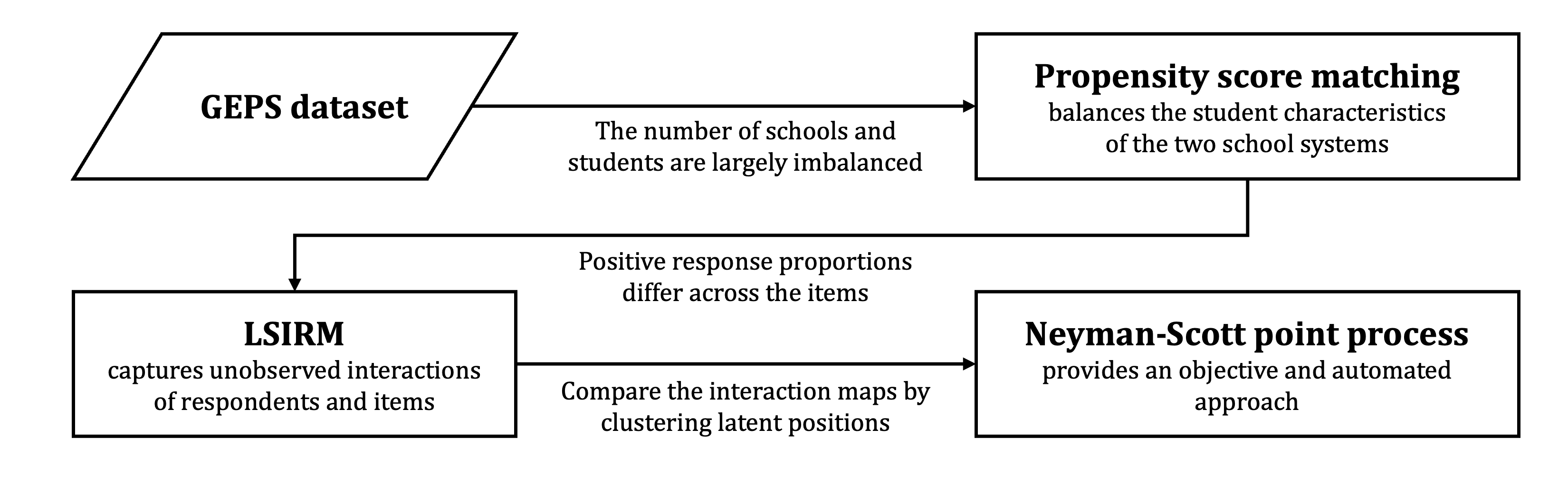}
    \caption{\label{fig:flowchart} \textcolor{black}{Flowchart for Section 2.}}
\end{figure}

\textcolor{black}{For better understanding, we start the organization of this section in Figure~\ref{fig:flowchart}.}
\subsection{GEPS Data Description} \label{DataDescription}

Here we provide a detailed description of the 2009 Gyeonggi Education Panel Study (GEPS) data (https://www.gie.re.kr/eng/content/C0012-04.do). Our analysis focuses on sixth-grade elementary school students who had undergone six years of an innovation or a regular school program. We investigate the differences in the non-cognitive behaviors of the students that were developed from the two school systems. Among the total of 2,960 elementary school students, 645 students came from 17 innovation schools; the remaining 2,315 students were  from 54 regular schools in Gyeonggi province. 

To evaluate the non-cognitive outcomes of the students, we used 62 items from the GEPS questionnaire. 
\textcolor{black}{
In the questionnaire, the items were divided into eight categories, each designed to measure corresponding aspects of students' attitudes toward their lives and their non-cognitive behaviors: }
Mental Ill-being (items 1 -- 6), Sense of Citizenship (items 7 -- 22), Self-Efficacy (items 23 -- 30, 61), Disbelief in Growth (items 31 -- 33), Test Stress (items 34 -- 40), Relationship with Friends (items 41 -- 46), Self-Esteem (items 47 -- 50, 62), and Academic Stress (items 51 -- 60).  
\textcolor{black}{
Each category has a varying number of items, ranging from three to thirteen. Each item is measured using a five-point Likert scale consisting of five response categories: (1) `Not at all', (2) `Not really', (3) `Usually', (4) `Yes', and (5) `Very much'. We dichotomized the responses such that (4) `Yes' and (5) `Very much' indicate `positive' responses, while 
(1) `Not at all', (2) `Not really', and (3) `Usually'  indicate 'negative' responses. 
}
All items are presented in the Supplementary Material.

\begin{figure}[htbp]
\centering
\includegraphics[width=0.55\textwidth]{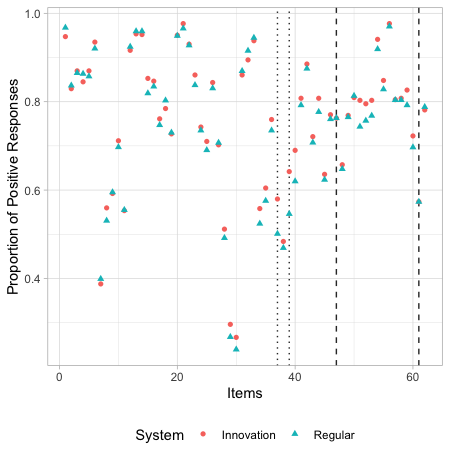}
\caption{\label{fig:item_score}
The proportion of positive responses for the 62 test items across all students from the regular and innovation schools. 
Dotted lines show items 37 and 39, and dashed lines show items 47 and 61. }
\end{figure}

Figure \ref{fig:item_score} displays the proportion of positive responses of all 62 items across the students from the regular and innovation school systems. The mean difference in the positive response proportion between the two school systems is 0.0184. 
However, the difference varies by item. Items 39 and 37 show the largest mean difference (0.0957 and 0.0787 each), belonging to the ``Test Stress'' category. This suggests that the two school systems may differ in their approach to student assessment and stress management.

On the other hand, Items 8 and 23 show the largest mean difference such that the regular school has higher positive response proportion. Each item belongs to ``Sense of Citizenship'' and ``Self-Efficacy'' category, implying that there exist difference in how students behave to society and to themselves. 
Items 61 (Self-Efficacy) and 47 (Self-Esteem) show the least difference between the systems, implying that both systems may have similar effects on the perceptions of students toward self-efficacy and self-esteem. 

{\color{black}{
Notably, the positive response proportions differ across the test items but also between the regular and innovation school systems. This implies that univariate analysis based on aggregated measures may not fully capture the complexity of the differences between the two school systems. To better understand these differences at the item level, we conduct a preliminary analysis using the latent space item response model (LSIRM) following \cite{jeon2021mapping}.}}

\subsection{Matching between Innovation and Regular Schools}\label{sec:match}

\textcolor{black}{Before proceeding with LSIRM analysis, we first match the data sets to balance the student characteristics of the two school systems.}
As noted in Section~\ref{DataDescription} the number of schools and students are largely imbalanced between innovation schools and regular schools. 
Also, students' personal backgrounds, such as parent support, may influence students' responses to the items. To examine differences that could be attributed to the two school systems accurately and minimize potential confounding, we match the student data from the two school systems with respect to socioeconomic status (SES).

We apply optimal propensity score matching \citep[PSM]{doi:10.1198/106186006X137047}. PSM is a widely used statistical technique  that aims to create comparison groups in observational studies with similar characteristics to reduce selection bias and enable causal inference \citep{doi:10.1080/00273171.2011.568786}. 
Students' SES is a major factor that shapes their perceptions about schools, mental well-being, and academic achievements \citep{aikens2008socioeconomic, Xuan2019, Baik2015}. Therefore, we selected SES indicators (15 items from school survey data and 21 items from parent survey data) and performed optimal PSM to match innovation and regular school students based on SES. 
For example, some of the SES indicators used for matching included the ratio of students attending school with support, the amount of budget the school received, and the relationship between students and their parents. The full SES item list and mean responses are provided in the Supplementary Material. We used the \texttt{optmatch} \texttt{R} package \citep{doi:10.1198/106186006X137047} for the matching process. 
As a result, we obtained a sample of 17 regular schools with 698 students that are comparable to the 17 innovation schools with 645 students. These matched samples were used 
in all subsequent analyses. 

\subsection{LSIRM}\label{LSIRM}

We first describe the LSIRM approach that we employ to examine the relationships between respondents and items in the regular and innovation schools. \textcolor{black}{We discuss the results and  the need for the NS point process approach.}


Let ${\bf X} \in \mathbb R^{N \times P}$ be a binary item response matrix where $N$ and $P$ denote the number of respondents (students) and items, respectively. Then, the probability of giving an affirmative response (e.g., ``agree'') to item $i$ by respondent $k$ in the latent space item response model (LSIRM; \citealp{jeon2021mapping}) is
\begin{equation}\label{eq:lsirm_likelihood}
\mbox{logit} \Big(P\big(X_{ki} = 1 \mid \theta_k, \beta_i, \gamma, {\bf z}_k, {\bf w}_i\big)\Big) = \theta_k + \beta_i - \gamma \cdot d\big(\mathbf{z}_k, \mathbf{w}_i\big),
\end{equation} 
where $\theta_k$ and $\beta_i$ are the main effects for the respondent $k$ and the item $i$, indicating the person's latent trait being measured and item's threshold, respectively.  Here ${\bf z}_k$ is the latent position of the respondent $k$, ${\bf w}_i$ is the latent position of the item $i$, $d(\mathbf{z}_k,  \mathbf{w}_i)$ is a $l_2$ distance (Euclidean distance) between ${\bf z}_k$ and ${\bf w}_i$, and $\gamma \geq 0$ is the weight of the distance term. Here we set $\gamma > 0$ to ensure 
the inverse relationship between the distance and the probability of a positive response; the closer the person and item latent positions ($\mathbf{z}_k$ and $\mathbf{w}_i$) are, the more likely a student $k$ gives a positive response to the item $i$, given the person's latent trait and the item's threshold levels. We assume that the latent space is defined over the $\mathbb R^2$ domain.

\paragraph*{Advantages}
Compared to the traditional item response models \citep{hulin1983item}, which assume item responses are independent for given the respondent and item attributes (i.e., conditional independence assumption), LSIRM allows for complex interactions between respondents and items. 
A unique merit of LSIRM is that it provides a geometric representation of the interactions among respondents and items in the shared metric space. The interaction structure mapped onto the two-dimensional Euclidean space reveals unobserved similarities and dissimilarities.
\textcolor{black}{
Specifically, LSIRM can explain interactions (1) between respondents from $\lbrace {\bf z}_1,\cdots, {\bf z}_N \rbrace$, (2) between items from $\lbrace {\bf w}_1, \cdots, {\bf w}_P \rbrace$, and (3) between respondents and items from $
\lbrace {\bf z}_1,\cdots, {\bf z}_N \rbrace$, $\lbrace {\bf w}_1, \cdots, {\bf w}_P \rbrace$. In terms of respondent-by-respondent relationships, the closer ${\bf z}_k$, ${\bf z}_{k'}$ are, the more likely respondents $k$ and $k'$ have similar response patterns. The closeness between ${\bf z}_k$ and ${\bf z}_{k'}$ are simply calculated from the Euclidean distance $\|{\bf z}_k-{\bf z}_{k'}\|^2$ defined over $\mathbb R^2$ domain. The other types of relationships can also be similarly explained. 
}
Therefore, inspecting the latent space configurations of the innovative and regular school systems could provide new insights into the differences in their students' behaviors at the item level  between the two school systems.

\paragraph*{Estimation}

To estimate parameters in LSIRM, we apply Bayesian inference with the Markov chain Monte Carlo (MCMC) method. We specified the prior distribution of parameters as
\[\begin{split}
    &\beta_i \sim N(0, \sigma_\beta^2), \qquad \theta_j \mid \sigma_\theta \sim N(0, \sigma_\theta^2), \qquad \mathbf{z}_k \sim \mbox{MVN}(\mathbf{0}, \mathbf{I_2}), \\
    &\mathbf{w}_i \sim \mbox{MVN}(\mathbf{0}, \mathbf{I_2}), \qquad \log{\gamma} \sim N(\mu_\gamma, \sigma_\gamma^2), \quad \mbox{and} \quad \sigma_\theta^2 \sim \mbox{Inv-Gamma}(a_\sigma, b_\sigma).
\end{split}\]

\textcolor{black}{
Following \cite{jeon2021mapping}, we set $\sigma_{\beta} = 1$ and $a_{\sigma} = b_{\sigma} = 0.001$ in our implementation. Note that $\gamma$ serves as a weight parameter for the interaction term $d(\bf{z}_k, \bf{w}_i)$, and can be interpreted as the standard deviation of the latent positions of items and respondents. This implies that $\gamma$ determines the size of the interaction map. Therefore, to make a valid comparison between the innovation and regular school system, we fix $\gamma=1$ to standardize the size of the interaction map into [0,1] size across both systems. To confirm that $\gamma$ is positive in both school systems, we ran LSIRM with a spike-and-slab prior for $\gamma$. The posterior estimates of $\gamma$ were around 1.5 for both school systems with 95\% credible intervals (1.45, 1.60), and inclusion probabilities of $\gamma$ were larger than 0.99. This not only shows the existence of interactions between items and respondents, but also justifies fixing $\gamma$ to a positive value. 
}

Then, the full joint posterior distribution of the parameters given response data in LSIRM can be given as follows:
\begin{equation}\label{eq:lsirm_posterior}
\begin{split}
    \pi(\beta, \theta, \mathbf{z}, \mathbf{w}, \gamma \mid {\bf X}) &\propto \prod_k \prod_i P({\bf X}_{ki} = {\bf x}_{ki} \mid \beta_i, \theta_k, \mathbf{z}_k, \mathbf{w}_i, \gamma)\\
    &\times \pi(\gamma) \prod_i \pi(\beta_i) \prod_k \pi(\theta_k) \prod_k\pi(\mathbf{z}_k) \prod_i \pi(\mathbf{w}_i). 
\end{split}
\end{equation}

Note that the likelihood function from \eqref{eq:lsirm_likelihood} is invariant to translations, reflections, and rotations of the positions of the respondents and items. 

\begin{figure}[htbp]
    \centering
    \includegraphics[width=0.4\textwidth]{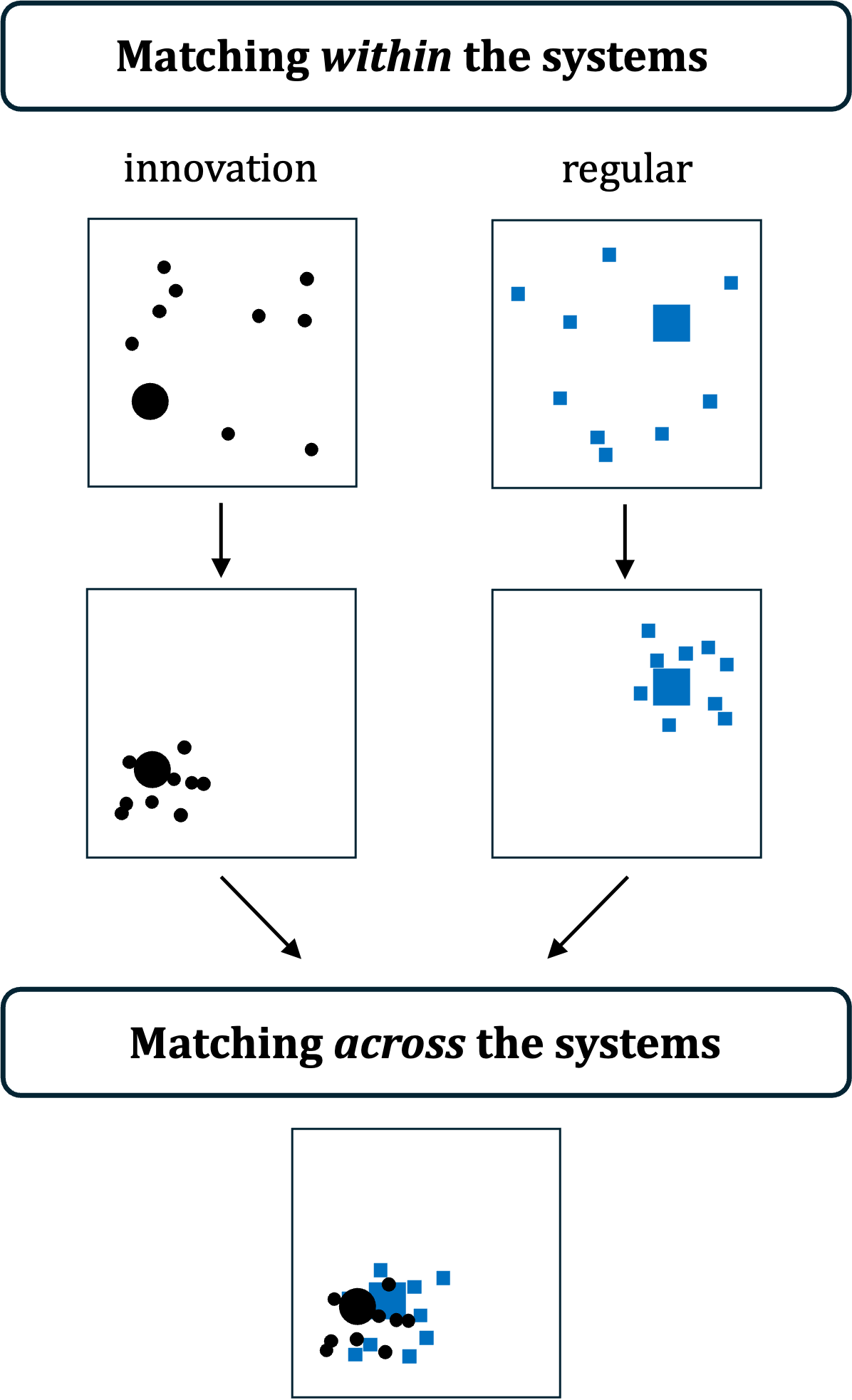}
    \caption{\textcolor{black}{Illustration of the two-step Procrustes matching procedure applied to MCMC samples of the latent positions $\mathbf{w}_i, \mathbf{z}_k$. The top four boxes represent the first matching process (matching within the systems). Large points represent the reference set within each system, and small points (MCMC samples) are matched to the large points. The bottom box shows the second matching process (matching across the systems). The aligned samples from the regular school system (represented by blue square points) are transformed to match the samples from the innovation school system (represented by black circle points).}}
    \label{fig:proc_example}
\end{figure}

\textcolor{black}{
To address this issue, we apply a post-processing using two-step Procrustes matching to align the Markov chain Monte Carlo (MCMC) samples within each school system and between the two school systems \citep{gower1975generalized}. After we obtain posterior samples from (2) through the MCMC algorithm, we implement Procrustes matching method \citep{friel2016interlocking} as follows. 
\begin{enumerate}
    \item[1.] Within each school system, we find a reference set of latent positions that achieves the highest value of the full log posterior density (represented by solid lines in Figure~\ref{fig:proc_example}). We then apply the first Procrustes matching to all other MCMC samples, aligning them to the reference set of each system. This step ensures consistency of latent positions within the systems.
    \item[2.] We conduct the second Procrustes matching across the school systems. MCMC samples from the regular school system (represented by the blue square points in Figure~\ref{fig:proc_example}) are transformed using Procrustes matching on the samples from the innovation school system (represented by the black circle points in Figure~\ref{fig:proc_example}). This adjustment minimizes the distance between the two sets of data, enabling direct comparison of their underlying structures.
\end{enumerate}
In the second matching process, MCMC samples of ${\bf w}_i$ from the regular school system are first transformed using Procrustes matching on the ${\bf w}_i$ samples from the innovation school system. Samples of ${\bf z}_k$, the latent position of the students, are then transformed using the same rotation matrix. The first step of Procrustes matching addresses the inherent uncertainties of LSIRM, while the second step facilitates a comparison by aligning the latent positions across the two school systems. 
}

In our application, we run MCMC algorithm for 60,000 iterations with the first 10,000 discarded for burn-in and 5,000 thinned samples are obtained from the remaining 50,000 for all parameters. 
\textcolor{black}{
Since the dimension of the parameters in LSIRM grows with $N$ and $P$, there can be potential memory issues without thinning. 
For model assessment, we provide the trace plots and R-hat values \cite{10.1214/ss/1177011136}) of the parameters in the Supplementary Material. We also performed posterior predictive checks by comparing the proportion of positive responses from the observed data and the fitted model. We observe that our method fits the observed data well (See the Supplementary Material for details).  
}

\begin{figure}[htbp]
\begin{center}
\begin{tabular}{cc}
\textbf{(a) Innovation} & \textbf{(b) Regular} \\
\includegraphics[width=0.45\textwidth]{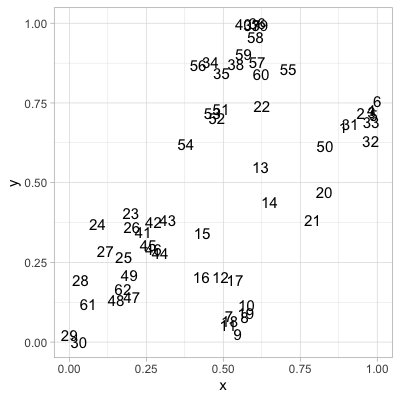} &
\includegraphics[width=0.45\textwidth]{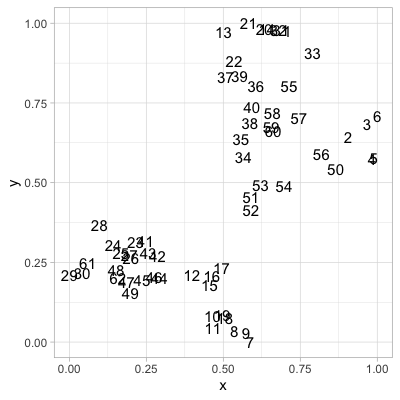}
\end{tabular}    
\end{center}
\caption{Item interaction map for innovation and regular schools. The numbers represent the posterior mean of the latent positions $\mathbf{w}_i$ from the latent space item response model. }
\label{fig:E_wproc} 
\end{figure}

\paragraph*{Preliminary Results}
{\color{black}{
We present interaction maps of the 62 survey items (Figure~\ref{fig:E_wproc}) obtained by fitting the LSIRM. The interaction maps offer a visual representation of the relationships among survey items and respondents. 
While the interaction maps provide valuable insights into the relationships between items and respondents, a direct comparison between the two interaction maps (innovation versus regular) is challenging. Clustering the latent positions could be a useful approach to facilitate the interpretation of the differences between the two school systems. Previous studies \citep{jin2022hierarchical} applied standard clustering methods such as spectral clustering for this purpose. However, such conventional methods have important limitations by requiring users to predetermine the number of clusters and not taking into account the uncertainties inherent in the clustering process. 

To address these limitations and provide an objective and automated approach to clustering, we propose the use of a Neyman-Scott point process model \citep{thomas1949generalization, neyman1952theory}. Neyman-Scott (NS) model allows us to identify clusters without specifying the number of clusters in advance and also accounts for uncertainties in the clustering process. In the next section, we discuss the NS model in more detail and explain how they are used to analyze the differences between the innovation and regular school system.}}

\subsection{Neyman-Scott Point Process Model}


To briefly describe the NS process modeling approach, consider $\mathbf{W}=\lbrace \mathbf{w}_1,\cdots,\mathbf{w}_P \rbrace$ be the latent positions of items obtained from the LSIRM on the bounded domain $\mathcal{S}\in \mathbb{R}^{2}$. In practice, we can use the posterior mean of latent positions from the LSIRM fit. Then we can define the NS process $\mathbf{W}$ as a collection of offsprings (clusters) $\bigcup_{c \in C} \mathbf{W}_c$, where $\mathbf{C}=\lbrace \mathbf{c}_1, \cdots, \mathbf{c}_{M} \rbrace$ is the parent (cluster center) of $\mathbf{W}_c$. Here, $\mathbf{C}$ follows a Poisson process with intensity parameter $\kappa$. Given the centers of the item group (parents) $\mathbf{C}$, the groups of items (offsprings) $\mathbf{W}_c, c \in \mathbf{C}$ follow an independent Poisson process as
\begin{equation}
p(\mathbf{W}|\mathbf{C},\alpha,\omega) = \exp\Big(|\mathcal{S}|-\int_\mathcal{S}\sum_{i=1}^{M} \alpha k(\mathbf{u}-\mathbf{c}_i,\omega)d\mathbf{u} \Big)\prod_{j=1}^{P}\sum_{i=1}^{M} \alpha k(\mathbf{w}_j-\mathbf{c}_i,\omega),
\label{Xlik}
\end{equation}
where $|\mathcal{S}|$ is the area of the interaction map domain. $k(\mathbf{u}-\mathbf{c}_i,\omega)$ is a Gaussian kernel with a center at $\mathbf{c}_i$ with variance $\omega^2$; the offspring are distributed around the parents. Here, $\alpha$ indicates the expected number of items for each group center $\mathbf{c}_i$ and $\omega$ controls the range of item groups in the interaction map.

\paragraph*{Estimation}

Although we have three parameters $(\kappa, \alpha, \omega)$, representing $\kappa$ in terms of $\alpha$ is preferable for accurate Bayesian inference \citep{kopecky2016bayesian}. Especially $\kappa = \frac{P}{|\mathcal{S}|\alpha}$, where $\frac{P}{|\mathcal{S}|}$ is an overall intensity of items in the interaction map domain. Then the joint posterior distribution is
\begin{equation}
\pi(\mathbf{C},\alpha,\omega|\mathbf{w}) \propto p(\mathbf{W}|\mathbf{C},\alpha,\omega)p(\mathbf{C}|\alpha)\pi(\alpha)\pi(\omega),
\label{PointprocessPost}
\end{equation}
where $\pi(\alpha),\pi(\omega)$ are priors of $\alpha$ and $\omega$, respectively. In the NS process, $\alpha$ and $\omega$ have a strong dependence on each other. With the same observed pattern, two possible interpretations are available; large $\alpha$ and large $\omega$ show a small number of parents with each having a large number of offsprings, whereas small $\alpha$ and small $\omega$ show the opposite. Therefore, to avoid such identifiability issues, we use uniform priors to set upper and lower bounds for $\alpha$ and $\omega$ \citep{moller2007modern, kopecky2016bayesian}. 
\textcolor{black}{
When setting the prior, we recommend incorporating the survey’s design and objectives for deriving more meaningful inference. In our study, we used the original eight categories from the GEPS questionnaire as a reference for the priors - especially, we set the range of $\alpha$ to $[\frac{62}{10|\mathcal{S}|}, \frac{62}{2|\mathcal{S}|}]$ so that we could obtain approximately 2 to 10 item clusters on average. For a sensitivity check, we explore different ranges of the $\alpha$ prior to compare our prior setting with both smaller and larger $\alpha$ ranges. As a guide for applied researchers, we recommend setting the prior range based on the prior knowledge on the item characteristics or questionnaire categories. We suggest including $\frac{N}{m |\mathcal{S}|}$ within the prior range of $\alpha$, where $N$ denotes the total number of items, $\mathcal{S}$ represents the domain, and $m$ is the desired number of clusters. By setting the appropriate $\alpha$ range, researchers can obtain a desired number of item clusters that better aligns with the original survey design and research objectives. 
}

We use a positive non-informative uniform prior for $\omega$ because it is a range parameter. Given the posterior mean of the latent positions $\mathbf{W}$, we update \eqref{PointprocessPost} using an MCMC algorithm proposed in \cite{kopecky2016bayesian}. Our Bayesian approach updates $\mathbf{C}$ through a birth-to-death MCMC \citep{moller2003statistical} and the Metropolis-Hastings algorithm to update the model parameters $(\alpha, \omega)$. We provide details for the algorithm in the Supplementary Material.
 
\paragraph*{Clustering} To quantify the uncertainties in clustering, we repeat running an MCMC algorithm 1,000 times independently in our analysis. Since each MCMC run provides different $M$ (the number of clusters) and $\mathbf{C}$ (cluster centers), we can naturally obtain the distributions of $M$ and $\mathbf{C}$, respectively. The mode for the distribution of $M$ can be considered to be the {\it promising} number of clusters. For given a selected $M$ (i.e., mode), we chose the location of the optimal cluster centers $\mathbf{C}$ based on the Bayesian information criterion (BIC), which is calculated from the likelihood \eqref{Xlik}. We use the parent $\mathbf{C}$ with the minimum BIC as the optimal cluster center candidates.

Furthermore, we conduct contour plots-based adjustments to validate the $M$ cluster center candidates chosen above. From 1,000 independent NS process fittings, we can visualize the distribution of $\mathbf{C}$ via the contour plot. We identify any cluster centers that do not align with the peak of the contour plot. Especially if the cluster center is located within the low-density region of the contour plot, we remove the corresponding cluster center. Then we can obtain $M^{\ast}$ number of the adjusted cluster centers $\lbrace \mathbf{c}^{\ast}_{1}, \cdots, \mathbf{c}^{\ast}_{M^{\ast}} \rbrace$. Finally, we assign cluster membership to each of the items $\mathbf{w}_i$ based on the Euclidean distance as $\arg \min_{j} \| \mathbf{w}_i-\mathbf{c}^{\ast}_{j} \|$. By using both the distributions of $M$ and $\mathbf{C}$, we can quantify uncertainties in the clustering step.


\section{Analysis Results} \label{sec:results}

In this section, we analyze the GEPS data using the proposed method. 
We start with presenting the number of identified item clusters and their properties, followed by a comparison of the results between the two school systems. Next, we establish student clusters based on their relative location to the item clusters and interpret their features. Finally, we compare our item clustering results with a conventional approach, exploratory multidimensional item  response analysis.

\subsection{Item Map}\label{sec:itemmap}

\subsubsection{Item Clusters}

\begin{figure}[htbp]
\begin{center}
\begin{tabular}{cc}
\textbf{(a) Innovation} & \textbf{(b) Regular} \\
\includegraphics[width=0.45\textwidth]{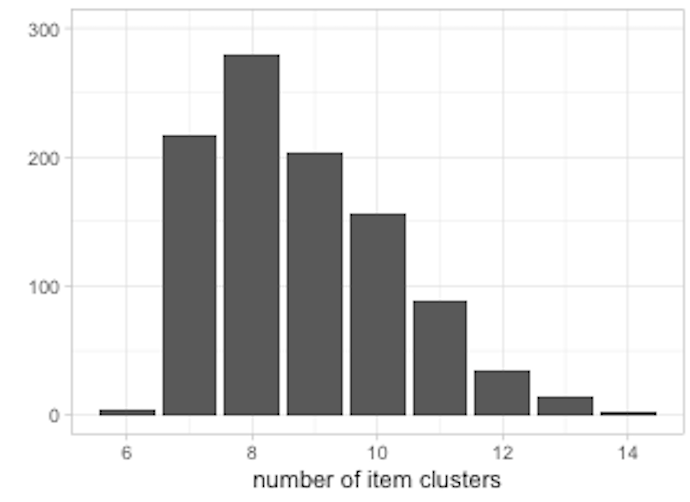} &
\includegraphics[width=0.45\textwidth]{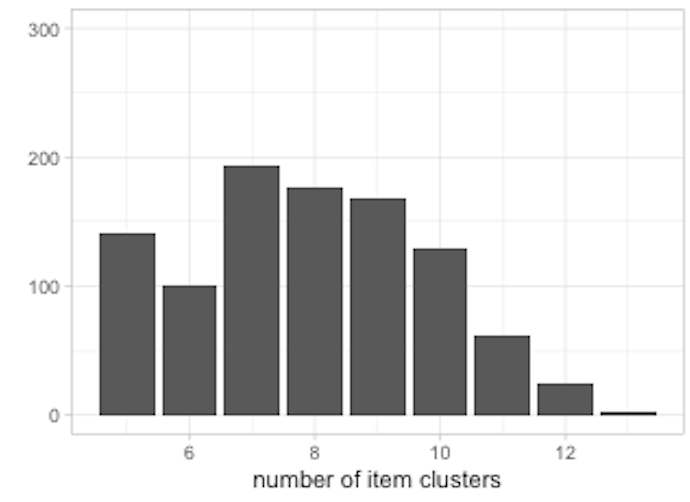}
\end{tabular}    
\end{center}
\caption{The distribution for the number of cluster centers obtained from 1,000 independent NS point processes.}
\label{fig:E_parent} 
\end{figure}

Figure~\ref{fig:E_parent} shows the distribution of the number of item clusters obtained from 1,000 independent NS point processes in the innovation and regular school systems. Based on the posterior mode in Figure~\ref{fig:E_parent}, we set the number of candidate clusters for the innovation and regular school systems at 8 and 7, respectively.

\begin{figure}[htbp]
\begin{center}
\begin{tabular}{cc}
\textbf{(a) Innovation} & \textbf{(b) Regular} \\
\includegraphics[width=0.45\textwidth]{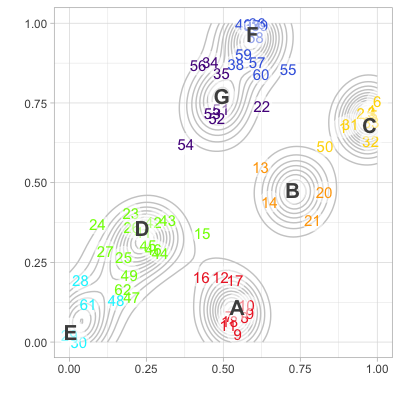} & 
\includegraphics[width=0.45\textwidth]{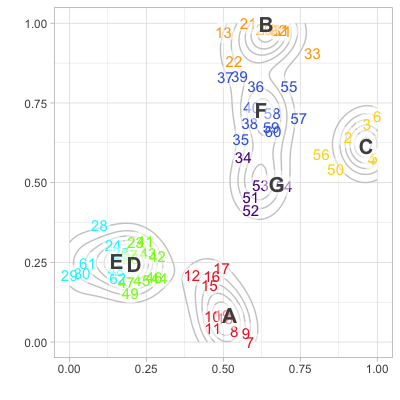}
\end{tabular}    
\end{center}
\caption{
Item interaction map for innovation and regular schools. Numbers represent item latent positions and alphabets represent item cluster centers, respectively. The grey lines are the distribution of the cluster centers obtained from 1,000 independent NS point processes. 
}
\label{fig:E_cluster}
\end{figure}

As we pointed out, Figure \ref{fig:E_parent}(a) suggests 8 clusters in the innovation school system, and we report the corresponding clustering results in the Supplementary Material (Figure 12 -- 15). However, based on the contour plot, cluster {\bf H} in the innovation school system does not have a peak, indicating that it may not be justifiable. Therefore, we adjusted the cluster centers in the innovation school system by removing cluster {\bf H} and obtained 7 cluster centers. The adjusted cluster centers and their corresponding items are shown in Figure \ref{fig:E_cluster}. From this, we can incorporate information on the distribution of cluster centers and quantify uncertainties in the clustering step. 
We will henceforth use the adjusted item interaction maps (Figure \ref{fig:E_cluster}) for further analysis.

\textcolor{black}{
As discussed in the previous section, the number of clusters depends on the prior settings for the $\alpha$ parameter. Figure 12 and 14 (Supplementary Material) show the distribution of the cluster numbers for smaller and larger $\alpha$ values. A smaller $\alpha$ range results in 9 and 12 clusters in innovation and regular school systems, leading to an over-separating items. On the other hand, a larger $\alpha$ range leads to 7 and 5 clusters in innovation and regular school system, merging items into fewer clusters. However, in this research, we focus on comparing the two school systems based on the original questionnaire design, therefore we use the results in Figure~\ref{fig:E_cluster}.
}

\subsubsection{Item Cluster Interpretation}

Figures \ref{fig:E_cluster}(a) and (b) display the interaction maps of items from the innovation and regular school systems, respectively. We observe that the item cluster centers tend to be located at the upper-right and lower-left sides of the map, suggesting that there are two principal directions for the underlying response patterns. As described in Section~\ref{LSIRM}, we assign cluster membership to each item based on the Euclidean distance between items and cluster centers. To directly compare clusters between the two school systems, we named clusters based on the properties of the grouped items. For example, cluster {\bf A} in both schools contains common items, such as items 7 -- 12 and 16 -- 19. 
Note that cluster {\bf A} in the innovation school does not have the same items as cluster {\bf A} in the regular school but has similar items; this naming convention is a simple and effective way to label the clusters based on their content. Based on the contents of the clustered items, we categorized 62 items into ten distinct properties: Sense of citizenship, Citizenship being conscious of others, Mental stability, Belief-in-growth, Sociability, Self-esteem, Self-confidence, Appearance-confidence, Test and academic confidence, and Test and academic confidence in terms of sociability. Each cluster contains one or more properties, and Table \ref{table:E_cluster} summarizes these results. By considering the properties of the items grouped, we can provide an interpretation of item clusters. 
\textcolor{black}{Note that the identified item clusters are from the LSIRM-NS approach, while the item categories (Section~\ref{DataDescription}) are from the survey design before we analyze the dataset. 
}

Cluster {\bf A} has ``The sense of citizenship'' property, which refers to the moral efforts of the students to live harmoniously in society. For example, items 9 (``I want to help a friend or neighbor who confronts difficulties around me'') and 16 (``I think a good society can be made by the efforts of citizens'') are included in cluster {\bf A}. Similarly, cluster {\bf B} also contains ``The sense of citizenship'' but differs from cluster {\bf A} in that it focuses on whether students are aware of other people when engaging in civic-conscious behavior. In particular, cluster {\bf B} includes item 13 (``I do not have to keep the law unless someone sees it'' (reverse-coded)) and 14 (``Although I witness a bad thing, I ignore it if it will not harm me'' (reverse-coded)). 

\begin{table}[htbp]
\centering
\begin{tabular}{c|c|c}
School system & Cluster & Property \\  \hline
\multirow{7}{*}{Innovation} 
 & A & \multicolumn{1}{l}{The sense of citizenship} \\
 & B & \multicolumn{1}{l}{Citizenship being conscious of others*} \\
 & C & \multicolumn{1}{l}{Mental stability*, Belief-in-growth*} \\
 & D & \multicolumn{1}{l}{Sociability, Self-esteem, Self-confidence}\\
 & E & \multicolumn{1}{l}{Appearance-confidence} \\
 & F & \multicolumn{1}{l}{Test and academic confidence*} \\
 & G & \multicolumn{1}{l}{Test and academic confidence in terms of sociability*} \\ \hline
\multirow{7}{*}{Regular}
 & A & \multicolumn{1}{l}{The sense of citizenship}\\
 & B & \multicolumn{1}{l}{Citizenship being conscious of others*, Belief-in-growth*} \\
 & C & \multicolumn{1}{l}{Mental stability*} \\
 & D & \multicolumn{1}{l}{Sociability, Self-esteem, Self-confidence} \\
 & E & \multicolumn{1}{l}{Sociability, Self-confidence, Appearance-confidence} \\
 & F & \multicolumn{1}{l}{Test and academic confidence*} \\
 & G & \multicolumn{1}{l}{Test and academic confidence in terms of sociability*} 
\end{tabular}
\caption{Item cluster property in innovation (top) and regular (bottom) school systems. Asterisk symbols indicate that the corresponding items are reverse-coded. 
}
\label{table:E_cluster}
\end{table}

Cluster \textbf{C} is characterized by ``Mental stability'', which reflects whether students have a positive attitude toward their lives. For instance, items 1 (``I am not interested in everything'' (reverse-coded)) and 2 (``I am worried about everything'' (reverse-coded)) are two examples of cluster {\bf C}, indicating how positive students are about their lives and, therefore, how stable their minds are. ``Belief-in-growth'' is a property that explains the need for efforts to improve one's ability. For example, items 31 (``A person’s ability is determined from the time of birth'' (reverse-coded)) and 33 (``Even if I try, my ability does not change much'' (reverse-coded)) demonstrate the property of belief-in-growth. Interestingly, belief-in-growth appears in cluster {\bf C} in the innovation school system, while it appears in cluster {\bf B} in the regular school system. This suggests that students in the innovation school system consider their belief-in-growth to be related to mental stability, while those in the regular school system believe it is associated more with their sense of citizenship being aware of others.

Cluster {\bf D} in both school systems is characterized by ``Sociability, Self-esteem, Self-confidence'', which indicates that students have high confidence in their social skills and strongly believe in their value. For example, students who responded ``yes'' to items 24 (``Friends like to play with me'') and 25 (``I am good at group activities'') are confident in their social skills. Moreover, items 47 (``I believe that I am as valuable as someone else'') and 49 (``I can do as well as someone else'') demonstrate the positive perception of the students about their worth.

Students who have answered ``yes'' to items grouped as ``Appearance-confidence'' are confident in their appearance above all else. For example, items 29 (``I have a favorable face'') and 30 (``My appearance is attractive'') illustrate this property well. However, in the innovation school, cluster {\bf E} consists only of items for the appearance-confidence property. On the other hand, in the regular school system, the appearance-confidence property appears together with sociability and self-confidence as cluster {\bf E}.

Clusters {\bf F} and {\bf G} exhibit differences with respect to the relationship between test and academic confidence and sociability. Cluster {\bf F} comprises items suggesting students are not anxious about exams or grades, such as items 40 (``I am just worried if it’s a test'' (reverse-coded)) and 60 (``It is burdensome for me to evaluate my ability with grades'' (reverse-coded)). In contrast, items in cluster {\bf G} are intended to assess the importance of friendship over academic achievement, as evidenced by items 53 (``I hate to show my preparation note for the college entrance exam to others'' (reverse-coded)) and 56 (``I can ignore friendship to get better grades in grades or entrance exam'' (reverse-coded)). These items differ significantly from those of cluster {\bf F}, as the latter considers the connection with friends, while the former does not.

\subsubsection{Item Cluster Comparison Between the Two School Systems}

\begin{table}[htbp]
\centering
\resizebox{\textwidth}{!}{%
\begin{tabular}{cccccccc|ccccccc}
 & \multicolumn{7}{c|}{\textbf{Innovation}} & \multicolumn{7}{c}{\textbf{Regular}} \\ 
 & \textbf{A} & \textbf{B} & \textbf{C} & \textbf{D} & \textbf{E} & \textbf{F} & \textbf{G} 
 & \textbf{A} & \textbf{B} & \textbf{C} & \textbf{D} & \textbf{E} & \textbf{F} & \textbf{G} \\ \hline
\textbf{A} 
&  & \textbf{0.41} & {0.71} & {0.40} & {0.55} & {0.86} & {0.66} 
&  & \textbf{0.92} & {0.69} & {0.35} & {0.40} & {0.65} & {0.44} \\
\textbf{B} & 
\textbf{0.41} &  & {0.32} & \textbf{0.50} & {0.85} & {0.51} & {0.38} & 
\textbf{0.92} &  & {0.50} & \textbf{0.87} & {0.89} & {0.27} & {0.50} \\
\textbf{C} & 
{0.71} & {0.32} &  & {0.81} & {1.17} & {0.48} & {0.49} & 
{0.69} & {0.50} &  & {0.84} & {0.89} & {0.36} & {0.31} \\
\textbf{D} & 
{0.40} & \textbf{0.50} & {0.81} &  & \textbf{0.40} & {0.71} & {0.49} & 
{0.35} & \textbf{0.87} & {0.84} &  & \textbf{0.06} & {0.64} & {0.53} \\
\textbf{E} & 
0.55 & 0.85 & 1.17 & \textbf{0.40} &  & \textbf{1.10} & \textbf{0.89} & 
0.40 & 0.89 & 0.89 & \textbf{0.06} &  & \textbf{0.67} & \textbf{0.57} \\
\textbf{F} & 
0.86 & 0.51 & 0.48 & 0.71 & \textbf{1.10} &  & 0.22 & 
0.65 & 0.27 & 0.36 & 0.64 & \textbf{0.67} &  & 0.24 \\
\textbf{G} & 
0.66 & 0.38 & 0.49 & 0.49 & \textbf{0.89} & 0.22 &  & 
0.44 & 0.50 & 0.31 & 0.53 & \textbf{0.57} & 0.24 &  \\ 
\end{tabular}
}
\caption{The Euclidean distance between the item cluster centers in innovation (left) and regular (right) school systems. Bold numbers indicate the case where the difference between the Euclidean distances of two school systems from the corresponding distances is greater than 0.3. }
\label{table:E_dist}
\end{table}

Table \ref{table:E_dist} presents the Euclidean distances between the centers of the item groups in each latent space of the school system. Figure~\ref{fig:E_cluster} and Table \ref{table:E_dist} indicate that clusters {\bf B} and {\bf E} exhibit distinct positions between the innovation and the regular school systems. Cluster {\bf B} is located closer to clusters {\bf A} and {\bf D} in the innovation school system than in the regular school system, as illustrated in Table \ref{table:E_dist} and Figures \ref{fig:E_cluster}. The proximity of clusters {\bf A} and {\bf B} in the innovation school system suggests that students in the innovation school system tend to help their neighbors and comply with the law regardless of others' awareness, while students in the regular school system have varied attitudes toward helping others based on the attention of their surroundings. Moreover, the adjacency of cluster {\bf B} to cluster {\bf D} in the innovation school system indicates a positive relation between civic behavior and self-confidence in one's social skills and self-worth. 
To compare the differences between the innovation and regular school systems, we analyzed the items from clusters {\bf A}, {\bf B}, and {\bf D} in detail. For example, on item 7 (cluster {\bf A}; ``I will actively participate in volunteer activities''), we found that among those who answered `yes' to this item, a higher proportion of students in the innovation school system (91.60\%) had positive responses to item 13 (cluster {\bf B}; ``I do not have to keep the law unless someone sees it'' (reverse-coded)) compared to the regular school system (84.72\%). Similarly, for item 23 (cluster {\bf D}; ``I like to hang out with others''), 4.40\% of the students within the same subgroup in the innovation school system had positive responses compared to 1.74\% of those in the regular school system. These findings suggest that students' attitudes toward citizenship differ between the two school systems, with students in the innovation school system more likely to comply with the law and enjoy getting along with others than those in the regular school system.

Table \ref{table:E_dist} and Figure \ref{fig:E_cluster} indicate that the relative position of cluster {\bf E} to clusters {\bf D}, {\bf F}, and {\bf G} is also significantly different between two school systems. In the innovation school system, cluster {\bf E} is located further away from clusters {\bf D}, {\bf F}, and {\bf G} than in the regular school system. This suggests that students in the regular school system tend to associate their appearance confidence with their social ability, while students in the innovation school system do not.

We also investigate the items from clusters {\bf D}, {\bf E}, {\bf F}, and {\bf G} in detail. We observe that among students who answered ``yes'' to item 29 (cluster {\bf E}; ``I have a favorable face''), a higher proportion of regular school students had positive responses to item 43 (cluster {\bf D}; ``I learn a lot from my school friends'') (86.24\% versus 81.15\%) and item 36 (cluster {\bf F}; ``When I take the exam, I forget what I know'' (reverse-coded)) (82.11\% versus 79.58\%) compared to innovation school students. These results indicate that regular school students tend to link appearance with their social and academic performance more than innovation school students. 


\subsection{Student Map}
\textcolor{black}{Section~\ref{sec:itemmap} explored the clustering of items based on their latent positions to uncover similarities in item properties. Here, we move our focus toward understanding student behaviors by clustering students according to their proximity to the item clusters identified in the previous section. 
By examining the students' positions relative to item clusters on the interaction map, we can infer which behaviors or tendencies (implied by the item properties) are more likely to resonate with students. }

\subsubsection{Student Clusters}

\begin{figure}[htbp]
\begin{center}
\begin{tabular}{cc}
\textbf{(a) Innovation} & \textbf{(b) Regular} \\
\includegraphics[width=0.45\textwidth]{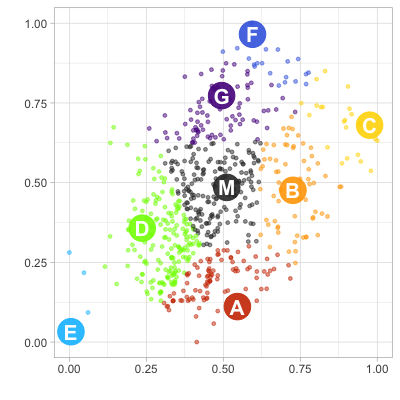} &
\includegraphics[width=0.45\textwidth]{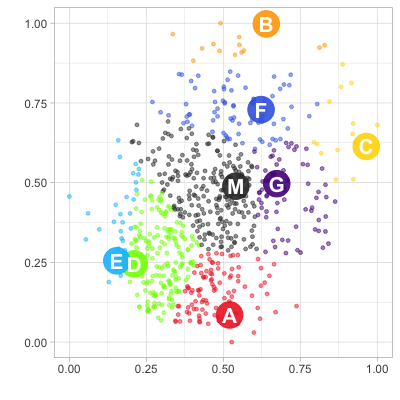}
\end{tabular}    
\end{center}
    \caption{Student interaction map for (A) innovation and (B) regular school systems. Dots represent student latent positions, and the colors show their cluster membership. }
    \label{fig:E_student}
\end{figure}

In Figure \ref{fig:E_student}, we overlap the students' latent positions, $\mathbf{z}_k$, over the item cluster centers from Section~\ref{sec:itemmap}. We obtain the posterior mean of $\mathbf{z}_k$ from LSIRM fitting and apply procrustes matching to visualize the interaction map. The proximity of the latent position of a student to an item cluster indicates their tendency to respond positively to items in that cluster. To further understand students' response behaviors, we assign each student to an item cluster based on the Euclidean distance between the student's latent position and the center of the item clusters.


We observe that most students are located at the center of the map, indicating they do not show strong preferences or tendencies toward specific item clusters. To distinguish students who exhibit stronger associations with specific item clusters from those who do not, we introduce a midpoint {\bf M}. \textcolor{black}{Note that M a midpoint of the identified cluster centers from Section~\ref{sec:itemmap} (rather than an identified cluster itself). We use M as a reference point to differentiate students who exhibit strong tendencies toward specific item clusters from those who do not show such preferences.}


\begin{table}[]
\centering
\begin{tabular}{c|cccccccc}
School system  & A   & B  & C  & D & E & F  & G  & M   \\ \hline
Innovation     & 93  & 66 & 23 & 176 & 3 & 20 & 81 & 183 \\
Regular & 99 & 13 & 14 & 175 & 31 & 77 & 69 & 220 \\ 
\end{tabular}
\caption{The number of students in each student cluster innovation (top) and regular (bottom) school systems. }
\label{table:E_Student}
\end{table}

\subsubsection{Student Cluster Comparison Between the Two School Systems}

Although the number of students in each cluster is generally similar in both systems, clusters {\bf B}, {\bf E}, and {\bf F} show significant differences. Clusters {\bf E} and {\bf F} in the innovation school system are located at the edge of the interaction map, resulting in fewer students being included in these clusters. In contrast, the number of students in cluster {\bf B} is higher in the innovation school system (66) than in the regular school system (13). This implies that the students in the innovation school tend to exhibit prosocial behaviors such as volunteering and complying with rules and regulations.

As pointed out in Section\ref{sec:itemmap}, item cluster {\bf B} is closer to cluster {\bf A} and {\bf D} in the innovation school system, implying a positive correlation between civic behavior and self-confidence in social skills for students in the innovation school system. Among students in cluster {\bf B}, 68.79\% and 20.98\% responded positively to the items in cluster {\bf A} in the innovation and regular school systems, respectively. Meanwhile, 51.72\% and 32.17\% answered positively to the items in cluster {\bf D}. The findings indicate that the student's attitudes toward citizenship vary depending on the school system they belong to, which in turn influences their views on sociability and self-confidence.


Students in cluster {\bf E} demonstrate notable differences between the innovation and regular school systems, particularly in their response patterns to items in clusters {\bf D}, {\bf F}, and {\bf G}. The students in cluster {\bf E} responded positively to the items in cluster {\bf D} with an average of 28.88\% and 86.80\% for innovation and regular schools, respectively. Such differences imply the contrasting ways in which students in the two school systems perceive and relate their confidence in appearance and social skills. Furthermore, all three students in cluster {\bf E} from the innovation school system responded negatively to the items in clusters {\bf F} and {\bf G}, indicating that there is no association between their confidence in test and academic performance and their sociability.


\subsection{Comparison with Multidimensional Item Response Theory Model}

One may think conventional methods, such as factor analysis, can also be applied to identify item clustering. Although our approach does more than item clustering and the goals of our analysis goes beyond identifying item clusters, it may be useful to provide some comparison with existing methods. Thus, we here provide a comparison with a conventional method focusing on item clustering results. 

We chose a multidimensional item response theory (MIRT) model, i.e., factor analysis for binary item response data, for the comparison. We estimated the MIRT model using the \texttt{mirt} \texttt{R} package with oblique rotation \citep{JSSv048i06}. The number of factors was determined to be four based on BIC. The factor loadings and item membership for each school system are provided in the Supplementary Material. 

\begin{table}[htbp!]
\centering
\begin{tabular}{c|c|l}
\textbf{School system} & \textbf{Factor} & \multicolumn{1}{c}{\textbf{Property}} \\ \hline
\multirow{4}{*}{Innovation} & 1 & Self-confidence, Sociability, Self-esteem \\ \cline{2-3} 
 & 2 & \begin{tabular}[c]{@{}l@{}}Mental stability, Sense of citizenship, Belief-in-growth, \\ Academic confidence, Sociability (Neg.)\end{tabular} \\ \cline{2-3} 
 & 3 & Test and academic confidence, Sociability \\ \cline{2-3} 
 & 4 & Sense of citizenship, Sociability, Self-esteem \\ \hline
\multirow{4}{*}{Regular} & 1 & Self-confidence, Sociability, Self-esteem \\ \cline{2-3} 
 & 2 & \begin{tabular}[c]{@{}l@{}}Mental stability, Sense of citizenship, Belief-in-growth, \\ Academic confidence, Self-esteem (Neg.)\end{tabular} \\ \cline{2-3} 
 & 3 & Test and academic confidence \\ \cline{2-3} 
 & 4 & \begin{tabular}[c]{@{}l@{}}Sense of citizenship, Sociability, Self-esteem, \\ Academic confidence, Test confidence (Neg.)\end{tabular}
\end{tabular}
\caption{Factor structures in the innovation (top) and regular (bottom) school systems. (Neg.) indicates that the corresponding items have negative loadings ($<0$). 
}
\label{table:E_factor}
\end{table}

In Table \ref{table:E_factor}, we observe that the properties of each factor are similar across both school systems. Factor 1 consists of items related to self-confidence, sociability, and self-esteem. In particular, items measuring self-confidence and sociability (items 23 -- 30) have high loadings ($>0.6$) in both school systems. However, in the regular school system, items measuring sociability (items 41 -- 46) have higher loadings than in the innovation school system. Factor 2 includes properties of mental stability, the sense of citizenship being conscious of others, belief-in-growth, and academic confidence. Items measuring mental stability have high loadings in both school systems (items 3 -- 6), indicating that they are the main property of the factor. However, it is important to note that the sociability property (items 41 -- 46) has negative loadings in the innovation school system, and the self-esteem property (item 62) has negative loadings in the regular school system. This suggests that students with stable mental health tend to be less social in the innovation school system and have lower self-esteem in the regular school system. 
Factor 3 is related to the test and academic confidence. Although factor 3 includes the property of sociability (items 41 -- 46) only in the innovation school system, these items have small loadings. In both school systems, seven items measuring test confidence (items 34 -- 40) have high loadings, being the unique feature of factor 3. 
Finally, factor 4 reflects a sense of citizenship and sociability. Although both school systems have items with negative loadings (e.g., item 22 in the innovation school system and items 13 and 37 in the regular school system), these items do not load as high as the positive items. The regular school system includes the properties of test and academic confidence; however, these items are not unique to factor 4 and do not load as highly on this factor. Therefore, factor 4 reflects primarily a sense of citizenship and sociability.

Compared to the clustering results from LSIRM-NS approach, factor 1 corresponds to clusters {\bf D} and {\bf E}. Factor 2 relates to clusters {\bf B}, {\bf C}, and {\bf D} in the innovation school system and to clusters {\bf B}, {\bf C}, and {\bf F} in the regular school system. Factor 3 refers to clusters {\bf F} and {\bf G} (also cluster {\bf D} in the innovation school system), and factor 4 covers clusters {\bf A} and {\bf D} (in addition, cluster {\bf G} in the regular school system). Compared to the results from MIRT, our method provides a larger number of clusters, providing different perspectives. For example, our model distinguishes between clusters {\bf D} and {\bf E} based on appearance-based confidence, whereas factor 1 in the MIRT model combines all items related to confidence without distinction. 

Although both our LSIRM-NS model and  MIRT can produce item clustering, 
our LSIRM-NS approach has several advantages over MIRT as follows:  
First, our method can provide more detailed understanding of the item response datasets by capturing the latent positions of both items and students. Especially our method can reveal various types of relationships between items, between respondents, and between items and respondents, while MIRT cannot. 
Moreover, the MIRT method may assign items to multiple factors, making it challenging to interpret the properties of each factor. In contrast, our method utilizes the NS process to cluster items based on their latent positions, ensuring each item is assigned to a single cluster.
Lastly, the LSIRM-NS approach can quantify uncertainties in both estimation and clustering steps. In the estimation step, we can obtain the posterior distribution of the latent positions of items and respondents rather than providing point estimates only. In the clustering step, the number and location of cluster centers are automatically determined from their distributions. On the other hand, for the MIRT approach, the number of clusters must be predetermined by users without accounting for uncertainties.

\section{Conclusion}

In this manuscript, we propose a novel framework to investigate the impact of the innovation school system on students' non-cognitive outcomes. By applying LSIRM to the GEPS questionnaire, we can simultaneously understand complex interactions between respondents and items. Furthermore, the NS point process fitting to latent positions allows us to automate the clustering procedure; therefore, we can provide a more objective interpretation of the latent space configuration, which is an important contribution.
\textcolor{black}{
Compared with conventional methods (e.g., MIRT), our procedure can quantify uncertainties that may be present in the process of determining clusters. Specifically, from the NS model fitting, we can obtain the  distribution of the number of clusters and contour plots of cluster centers; both are crucial for the reliable identification of clusters in the latent space.
}

Here, we study how the school system can form student attitudes and behaviors. In particular, students in the innovation school system tend to exhibit a stronger sense of citizenship, even when others are unaware of their actions. This behavior is also closely related to their self-confidence in social skills and self-worth. In addition, our findings suggest that students in the innovation school system are less likely to relate confidence in their appearance to confidence in their ability to socialize. In contrast, this tendency is more prevalent among students in the regular school system. Our method can be useful for policymakers in designing effective education systems by providing a comprehensive understanding of the complex relationships between students and the school systems. 

\textcolor{black}{
One may wonder whether the LSIRM with the NS model can be integrated into a single-step approach. While it is possible in principle, such an integration is likely to greatly increase uncertainty of latent position estimation (due to the invariance property of latent positions) as well as computational cost (due to the necessity for matching within MCMC iteration). 
To control such unnecessary uncertainty and manage computational burden, we stay with a two-stage approach with Procrustes matching between the stages.
}

\textcolor{black}{
Our method has been made available in the {\tt R} package {\tt lsirm12pl}. 
However, there are some practical limitations to be considered in using the method. 
First, the dimension of the LSIRM model parameters increases with $N$ (number of respondents) and $P$ (number of items). Therefore, we recommend thinning posterior samples from LSIRM to prevent potential memory overload. Furthermore, in the independent NS process fitting step, we recommend parallel computation to accelerate the performance of the algorithm. In our analysis, we used  parallel computation through {\tt OpenMP} library in {\tt C++}, and the entire procedure took approximately 2.5 hours. Given that our data has a pretty large sample size, the proposed LSIRM-NS method is applicable to many realistic applications. Lastly, the priors for the NS model should be carefully specified to avoid identifiability issues \citep{moller2007modern, park2022interaction}. As discussed in Section~\ref{sec:itemmap}, researchers can use the designs and objectives of the survey questionnaires in use to construct informative priors. This approach ensures a realistic analysis of item response datasets across various disciplines. 
}

As per the future directions, one might consider constructing a formal statistical test to compare different interaction maps. Although our method can examine differences between the student responses from the two school systems, a rigorous testing method could be useful. Another possibility is to develop variants of LSIRM that can analyze the entire and subsets of the data simultaneously. While our approach provides a comprehensive understanding of the differences in students' non-cognitive behaviors, studying individual non-cognitive behaviors or subgroups of students is also important.

\section*{Acknowledgement}
This study was supported by the National Research Foundation of Korea (NRF-2020R1C1C1A0100386813, NRF-2020R1A2C1A01009881, RS-2023-00217705, and RS-2024-00333701) and ICAN (ICT Challenge and Advanced Network of HRD) support program (RS-2023-00259934) supervised by the IITP (Institute for Information \& Communication Technology Planning \& Evaluation). The authors are grateful to editor, associate editor, and anonymous reviewers for their careful reading and valuable comments. 

\bibliographystyle{Chicago}
\bibliography{Reference}


\end{document}